# OPEN STATISTICAL ISSUES IN PARTICLE PHYSICS[1]

By Louis Lyons

*Oxford University*

Many statistical issues arise in the analysis of Particle Physics experiments. We give a brief introduction to Particle Physics, before describing the techniques used by Particle Physicists for dealing with statistical problems, and also some of the open statistical questions.

**1. Introduction.** Particle Physics tries to delve into the structure of matter at its most basic level. It continues a tradition that dates back to the Greeks[2] or even earlier. In the early days of Chemistry, the smallest entities were atoms. Early in the 20th century, the experiments of Rutherford demonstrated that atoms consisted of a small nucleus, with the electrons circulating at distances of $\sim 10^{-10}$ metres. Subsequently, the nucleus was found to be made of protons and neutrons. Many other particles (known as hadrons) like protons and neutrons have subsequently been discovered, but within the last 30 years, the quark model has brought understanding to the multitude of what used to be called "elementary particles."

The entities that we currently believe are fundamental (i.e., they do not seem to have any sub-structure) are the quarks and leptons shown in Table 1. There are 6 of each, and they appear to be arranged in 3 "generations" of increasing mass, each containing quarks of electric charge $+2/3$ and $-1/3$ (in units where the electron's charge is $-1$) and leptons of charge $-1$ and 0. The neutral leptons are called neutrinos. Although charged leptons and neutrinos have been detected, quarks are believed to be confined within

Received September 2007; revised January 2008.

[1]Supported in part by a Leverhulme Foundation grant.

*Key words and phrases.* Particle Physics, parameter determination, goodness of fit, *p*-values, hypothesis testing, nuisance parameters, upper limits, blind analysis, signal-background separation, combining results.

[2]Although the notion of what constitutes a satisfactory theory has changed over the centuries, it has always been considered desirable that the number of basic elements out of which everything is constructed should number at most "**A FEW**." Since for the Greeks the basic elements were **A**ir, **F**ire, **E**arth and **W**ater, it is clear that they not only understood the basic principles of Science, but also had an excellent command of the English language.

Table 1
*The basic particles*

| Particle, charge | Generations | | |
|---|---|---|---|
| | 1 | 2 | 3 |
| Quark, $+2/3$ | u (0.3) | c (1.5) | t (175) |
| Quark, $-1/3$ | d (0.3) | s (0.5) | b (5) |
| Neutrino, 0 | $\nu_e(< 3*10^{-9})$ | $\nu_\mu(< 2*10^{-4})$ | $\nu_\tau(< 0.02)$ |
| Lepton, $-1$ | electron $(5*10^{-4})$ | $\mu(0.1)$ | $\tau(1.8)$ |

Masses shown in brackets are in $\text{GeV}/c^2$. In these units, the mass of the proton is 0.9.

hadrons. They have not been observed directly, but their existence is inferred from the simplification they bring to the multitude of hadrons, and to the way they explain many features of the way hadrons interact with each other or with leptons.

In addition to these particles, there are also others responsible for mediating the various fundamental forces. These include the massless photon $\gamma$, responsible for the electro-magnetic force; the massive $W$ and $Z$ bosons which mediate the weak force; and the gluons $g$ responsible for the strong force. In addition, there is the still to be detected graviton which mediates gravitational forces, and is usually denoted by the symbol    . Because the interacts so weakly it is hard to observe. Finally, there is the undiscovered Higgs boson, which is believed to be responsible for the mass of the other particles, and which is the object of intense searches in current experiments.

Of course, theoretical physicists are prolific at inventing models, and so there are many other suggested particles.

Experiments in Particle Physics are usually conducted at large accelerators, for example, at the European Centre for Nuclear Research (CERN) in Geneva, or at Fermi National Accelerator Lab near Chicago. CERN's soon-to-be-running Large Hadron Collider (LHC) is in a tunnel about 100 metres below the surface and 27 kilometres in circumference, and which straddles the French–Swiss border. Protons circulate in bunches in opposite directions around the ring, and collide with each other at the center of large detectors. The bunches are about the width of a human hair, and are ∼10 centimetres long. When they collide, new particles are produced by converting the available kinetic energy into mass. The detectors are designed to track the path of each particle, measure its curvature in the magnetic field and hence determine the particle's momentum, and also to give information on the particle's identity (e.g., whether it is an electron, muon, pion, kaon or proton).

Reactions between colliding protons will occur at a very high rate, but most of them are fairly uninteresting. Thus, experiments are designed to have a trigger, which makes a very fast decision as to whether the collision (called an "event") is likely to be interesting, and hence whether the data from the detector is worth storing. Because of data read-out and storage constraints, only about 100 events per second are recorded, and each may contain about a Megabyte of information. Since the accelerator may run for 15 years, some $10^{10}$ events can be collected by each experiment. In analyzing data, allowance must be made for the distorting effect introduced by any selection bias of the trigger.

This review attempts to present some interesting statistical issues in the analysis of data collected in Particle Physics experiments. The items discussed below are a mixture of current practice, ideals to which we aspire and some personal prejudices of the author. It is hoped that the approaches mentioned in this article will be interesting or outrageous enough to provoke some Statisticians either to collaborate with Particle Physicists, or to provide them with suggestions for improving their analyses. It is to be noted that the techniques described are simply those used by Particle Physicists; no claim is made that they are necessarily optimal.

A Glossary of Particle Physics terminology appears in the supplementary material [Lyons (2008)].

**2. Particle Physics analyses.** This section starts with two typical examples of Particle Physics analyses, the first involving parameter determination, while the second tests whether data is consistent with a null hypothesis $H_0$, or whether an alternative hypothesis $H_1$ is favored. Further examples are described later. More detailed descriptions can be found in the various papers of the PHYSTAT series of Conferences [see James, Lyons and Perrin (2000), Cheung and Lyons (2000), Whalley and Lyons (2002), Lyons, Mount and Reitmeyer (2003), Lyons and Ünel (2005), Reid, Linnemann and Lyons (2006), Prosper, Lyons and De Roeck (2007)]. In particular, at the PHYSTAT-LHC meeting at CERN in 2007, the major experiments at the LHC presented their statistical "wish-lists" [Gross (2007), Belikov (2007), Xie (2007)].

2.1. *Lifetimes.* Here we estimate the lifetime of some specific particle. Thus, we could have $n$ independent observations $t_1 \ldots t_i \ldots t_n$ for the times between the production and decay for this particle in the selected events. Then the mean lifetime $\tau$ could be determined by an unbinned likelihood fit to the probability density $\tau^{-1} \exp(-t/\tau)$. In real life we would have a more complicated expression, to allow for a possible background with a different time dependence, experimental resolution on the determination of $t_i$, and experimental acceptance of the detector and the trigger, which depends on $t$.

The various steps in the data analysis include:

- Reconstruct tracks from the hits in the detector.
- Select wanted events that are enriched in the particle whose lifetime we wish to measure.
- For each interaction, extract the decay time $t$ from $L$ and $v$, the distance the particle travels and its speed. Typical values are picoseconds, mms and 99% of the speed of light respectively.
- Model the signal, typically by an exponential time dependence probably smeared by time resolution effects, and the background. Time-dependent efficiencies for collecting the data may also be relevant.
- Perform a likelihood fit, to determine $\tau$ and its statistical error $\sigma_{\text{stat}}$.
- Estimate the systematic error $\sigma_{\text{syst}}$, and quote the result as $\tau \pm \sigma_{\text{stat}} \pm \sigma_{\text{syst}}$. These systematics [Heinrich and Lyons (2007)] can arise from uncertainties in some of the extra parameters involved in modeling the data (e.g., the level of background contaminating our signal), or from possible uncertainties in the theory (maybe the expected exponential decay distribution is complicated by the existence of two overlapping particles). Statisticians usually refer to the former as "nuisance parameters." In analyses involving enough data to achieve reasonable statistical accuracy, considerably more effort is devoted to assessing the systematic error than to determining the parameter of interest and its statistical error.
- Assess the goodness-of-fit between the data and the model, and ignore the estimated value for the parameter if the fit is unsatisfactory.

2.2. *Significant peak?* Another type of analysis might consist of looking at a mass spectrum (see Figure 1). In many situations we would expect to observe a rather smooth and somewhat boring distribution, but sometimes there may be a significant-looking peak at some mass position. This could correspond to the exciting discovery of a new particle, to a boring statistical fluctuation of the smooth background or to some unfortunately overlooked effect in the analysis.

We can make some numerical statement about the probability of obtaining a statistical fluctuation at least as extreme as the one we have observed. In this situation, we are performing a "Goodness of Fit" test, that is, we are comparing our data with the null hypothesis of a smooth distribution. Alternatively and probably more sensitively, we could use our data to compare the two hypotheses—just smooth background or an interesting peak above the background; this is "Hypothesis Testing."

2.3. *Bayes or frequentism?* In many analyses the question arises whether to use a Bayesian or a Neyman–Pearson Frequentist approach, or one which is neither (e.g., $\chi^2$, likelihood, etc.). Particle Physicists tend to favor a frequentist method. This is because we really do consider that our data are representative as samples drawn according to the model we are using (decay time distributions often are exponential; the counts in repeated time intervals do follow a Poisson distribution, etc.), and hence we want to use a statistical approach that allows the data "to speak for themselves," rather than our analysis being dominated by our assumptions and beliefs, as embodied in Bayesian priors. The reluctance to use priors is strongest in situations with several variables where multidimensional priors would be required, or in cases where very little is known about the relevant parameter—it may be acceptable to use prior information about a parameter which is already well measured, but more problematic to try to quantify prior ignorance.

However, in practice, it is very hard to use the Neyman frequentist construction when more than two or three parameters are involved: software to perform a Neyman construction efficiently in several dimensions would be most welcome. The choice of a useful ordering rule is also very important. Thus from a pragmatic point of view, even ardent frequentists are prepared to use Bayesian techniques. Most of them, however, would like to ensure that the technique they use provides parameter intervals with reasonable frequentist coverage. There are even mixed methods [Cousins and Highland (1992)] that use Bayesian priors for nuisance parameters, but a frequentist method for the parameter of interest. The thinking here is that, although such an approach cannot be justified from fundamentals, it provides a practical method whose properties can be checked, and are often satisfactory.

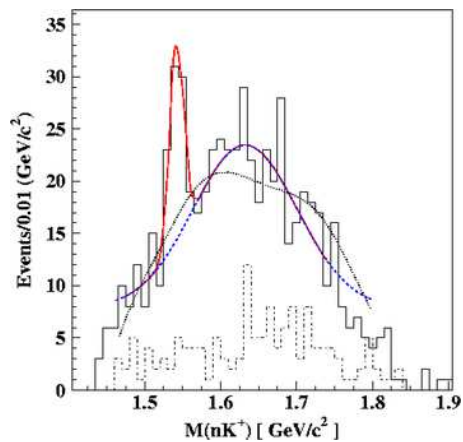

FIG. 1. *Mass histogram. This is for reactions producing a neutron ($n$), $\pi^+$, $K^+$ and $K^-$. A histogram of the effective mass of the $nK^+$ combination is plotted. If a particle decaying into a neutron and a $K^+$ is produced in these reactions, a narrow peak should appear in this histogram at the particle's mass, but if not the distribution should be smooth. The curve is an attempt to deduce this smooth background. Does the histogram provide evidence for a new particle, as opposed to there being a statistical fluctuation from the smooth background, and/or an incorrectly estimated background? A new particle here would be very interesting, as it would not fit into the simple quark model, because it would require a 5-quark structure.*

Particle Physicists would appreciate advice on how to construct priors for parameters of interest, to be used in conjunction with information-based priors for nuisance parameters, and which might give reasonable coverage [see Demortier (2005)].

**3. Experimental design.** Because experimental detectors are so expensive to construct, the time-scale over which they are built and operated is so long, and they have to operate under harsh radiation conditions, great care is devoted to their design and construction. This differs from the traditional statistical approach for the design of agricultural tests of different fertilisers, but instead starts with a list of physics issues which the experiment hopes to address. The idea is to design a detector which will provide answers to the physics questions, subject to the constraints imposed by the cost of the planned detectors, their physical and mechanical limitations, and perhaps also the limited available space. Inevitably, compromises in the design are required, and testing of any proposed scheme involves analysis of the simulated "data" to see if the physics aims can indeed be achieved.

Design is also involved when planning what technique is to be used to analyze the experiment's real data. This will be especially detailed if a blind analysis is to be performed (see Section 8).

Another example is provided by the attempt to assess the systematic error on an estimated parameter, caused by nuisance parameters. This often requires producing simulations of the data with different values of the nuisance parameter, and seeing how much the physics parameter's value changes when the nuisance parameter value is changed by its uncertainty (compare Sections 5.4 and 6.2 for ways of incorporating nuisance parameters in upper limit and in $p$-value calculations respectively). When several nuisance parameters are involved, there is the question of whether separate simulations should be produced, in each of which only one of the nuisance parameters is changed from its optimal value by its uncertainty; or whether it is better to generate simulations in each of which all nuisance parameters are simultaneously changed from their optimal values according to their expected (possibly correlated) multivariate distribution. The two methods are sometimes referred to unisim (or OFAT = **O**ne **F**actor **A**t a **T**ime) and multisim respectively. The question is which method requires less computing time to achieve the same accuracy for the systematic error [Roe (2007)].

How to assess systematics was much discussed at the Banff meeting [Reid, Linnemann and Lyons (2006)] and PHYSTAT-LHC [Read (2007), Neal (2007), Linnemann (2007)].

**4. Separating signal from background.** Almost every Particle Physics analysis uses some technique for separating signal from background. This is because only a fraction $f$ of the complete set of stored events (which because of the trigger can be a factor of $10^7$ down on the total reaction rate) will contain interactions of interest for the analysis being performed. Depending on the investigation being undertaken, $f$ could be as small as $10^{-8}$.

First some simple "cuts" are applied; these are generally loose selections on single variables, which are designed to remove background while barely reducing the signal. For example, the selected events could be required to have no more than 6 charged tracks. Then some more sophisticated analysis is performed, perhaps using more complicated derived variables, for example, the mass of a possible particle decaying into a kaon and 3 charged pions. To separate signal from background in the multi-dimensional space of the event observables, these analyses typically use methods like Fisher discriminants, boosted decision trees, artificial neural networks (including Bayesian nets), support vector machines, etc. [Prosper (2002), Friedman (2003, 2005)]. A description of the software available for implementing some of these techniques can be found in Narsky (2006) and Höcker (2007).

If a large data sample is available to perform an accurate measurement of a property of some particle, then it is not a disaster if there is some level of background in the finally selected events, provided that it can be accurately assessed and allowed for in the subsequent analysis. At the other extreme, the separation technique may be used to see if there is any evidence for the existence of some hypothesised particle (the potential signal), in the presence of background from well-known sources. Then the actual data may in fact contain no observable signal.

These techniques are usually "taught" to recognize signal and background by being given examples consisting of large numbers of events of each type. These may be produced by Monte Carlo simulation, but then there is a problem of trying to verify that the simulation is a sufficiently accurate representation of reality. It is better to use real data, but the difficulty then is to obtain sufficiently pure samples of background and signal. Indeed, for the search for a new particle, true data examples do not exist. However, it is the accurate representation of background that is likely to pose a more serious problem.

The way that, for example, neural networks are trained is to present the software with approximately equal numbers of signal and background events[3], and then to optimize the cost function $C$ for the network. This is defined as $C = \Sigma(z_i - t_i)^2$, where $z_i$ is the trained network's output for the $i$th event; $t_i$ is the target output, usually chosen as 1 for signal and zero for background; and the summation is over all testing events presented to the network. The problem with this is that $C$ is not what we really want to optimize. For a search for a new particle, this could be the sensitivity of the experimental upper limit in the absence of signal, while for an analysis measuring the properties (such as mass or lifetime) of some well-established particle, we would be interested in minimizing the error (including systematic effects) on the result.

---

[3]For searches for rare processes, it is clearly inappropriate to use the actual fractions expected in the data to determine the ratio of signal to background Monte Carlo events in the training sample, because the network could achieve an excellent score simply by classifying everything as background.

So the open questions are as follows:

- Is it possible to define what multivariate method will perform well in a given class of problems?
- How can we check that our multi-dimensional training samples for signal and background are reliable descriptions of reality?
- How many events are required for training?
- How should they be divided between signal and background, especially when there are several different sources of background?
- What is the best way of allowing for nuisance parameters in the models of the signal and/or background?
- Are there easy ways of optimizing on what is really of interest?

**5. Upper limits.**   Most searches for new phenomena have not found any evidence for exciting new physics. Recent examples from Particle Physics include searches for the Higgs boson, supersymmetric particles, dark matter, etc.; attempts to find substructure of quarks or leptons; looking for extra spatial dimensions; measuring the mass of a neutrino; etc. Rather than just saying that nothing was found, it is more useful to quote an upper limit on the sought-for effect, as this could be useful in ruling out some theories. An example of this was the experiment by Michelson and Morley in 1887 which attempted to measure the speed of the Earth with respect to the aether. No effect was seen, but the experiment was sensitive enough to lead to the demise of the aether theory.

A simple scenario is a counting experiment where a background $b$ is expected from conventional sources, together with the possibility of an interesting signal $s$. The number of counts $n$ observed is expected to be Poisson distributed with a mean $\mu = \epsilon * s + b$, where $\epsilon$ is a factor for converting the basic physics parameter $s$ into the number of signal events expected in our particular experiment; it thus allows for experimental inefficiency, the experiment's running time, etc. Then given a value of $n$ which is comparable to the expected background, what can we say about $s$? The true value of $s$ is constrained to be non-negative. The problem is interesting enough if $b$ and $\epsilon$ are known exactly; it becomes more complicated when only estimates with uncertainties $\sigma_b$ and $\sigma_\epsilon$ are available.

Even without the nuisance parameters, a variety of methods is available. These include likelihood, $\chi^2$, Bayesian with various priors for $s$, frequentist

Neyman constructions with a variety of ordering rules for $n$, and various *ad hoc* approaches. The methods give different upper limits for the same data.[4] A comparison of several methods can be found in Narsky ([2000](#)). The largest discrepancies arise when the observed $n$ is less than the expected background $b$, presumably because of a downward statistical fluctuation. The following different behaviors of the upper limit (when $n < b$) can be obtained:

- Frequentist methods can give **empty** intervals for $s$, that is, there are no values of $s$ for which the data are likely. Particle Physicists tend to be unhappy when their years of work result in an empty interval for the parameter of interest, and it is little consolation to hear that frequentist statisticians are satisfied with this feature, as it does not lead to under-coverage.

    When $n$ is not quite small enough to result in an empty interval, the upper limit might be **very small**. This could confuse people into thinking that the experiment was much more sensitive than it really was.[5]

- The Feldman–Cousins frequentist method [Feldman and Cousins ([1998](#))] that employs a likelihood-ratio ordering rule gives upper limits which **decrease** as $n$ gets smaller at constant $b$. (This can also occur in other frequentist methods.) A related effect is the growth of the limit as $b$ decreases at constant $n$. Thus, if no events are observed ($n = 0$), the upper limit for a 90% interval is 1.08 for $b = 3.0$, but 2.44 for $b = 0$. This is sometimes presented as a paradox, in that if a bright graduate student worked hard and discovered how to eliminate the expected background, they would be "rewarded" by obtaining a weaker upper limit.[6] An answer is that although the actual limit had increased, the sensitivity of the experiment with the smaller background was better. There are other situations—for example, various random choices of measuring instruments [Cox ([1958](#))]—where a measurement with better sensitivity can on occasion give a less-precise result.

- In the Bayesian approach, the dependence of the limit on $b$ is **weaker**. Indeed, when $n = 0$, the limit does not depend on $b$.

---

[4] By coincidence, the values obtained by the Bayesian approach with an (improper) flat prior for $s$ and by the Neyman construction for upper limits agree when $b = 0$.

[5] Bayesian methods that use priors with part of the probability density being a $\delta$-function at $s = 0$ can result in a posterior with an enhanced $\delta$-function at zero, such that the upper limit contains only the single point $s = 0$.

[6] The $n = 0$ situation is perhaps a special case, as the number of observed events cannot decrease as further selections are imposed to reduce the expected background. For nonzero observed events, if $n$ decreases with the tighter cuts (as expected for reduced background), the upper limit is likely to go down, in agreement with intuition. But if $n$ stays constant, that could be because the observed events contain signal, so it is perhaps not surprising that the upper limit increases.

- Sen et al. ([2008](#)) consider a related problem, of a physical non-negative parameter $\lambda$ producing a measurement $x$, which is distributed about $\lambda$ as a Gaussian of variance $\sigma^2$. As the observable $x$ becomes more and more negative, the upper limit on $\lambda$ **increases**, because it is deduced that $\sigma$ must in fact be larger than its originally quoted value.

In trying to assess which of the methods is best, one first needs a list of desirable properties. These include:

- Coverage: Even most Bayesian Particle Physicists would like the coverage of their intervals to match their quoted credibility, at least approximately. Because the data in counting experiments are discrete, it is impossible in any sensible way to achieve exact coverage for all $\mu$. However, it is not completely obvious that even Frequentists need coverage for every possible value of $\mu$, since different experiments will have different values of $b$ and of $\epsilon$. Thus, even for a constant value of the physical parameter $s$, different experiments will have different $\mu = \epsilon * s + b$. Thus, it would appear that, if coverage in some average (over $\mu$) sense were satisfactory, the frequentist requirement for intervals to contain the true value at the requisite rate would be maintained. This, however, is not the generally accepted view by Particle Physicists, who would like not to undercover for any $\mu$.

- Not too much overcoverage: Because coverage varies with $\mu$, for methods that aim not to undercover anywhere, some overcoverage is inevitable. This corresponds to having some upper limits which are high, and this leads to undesirable loss of power in rejecting alternative hypotheses about the parameter's value.

- Short and empty intervals: These can be obtained for certain values of the observable, without resulting in undercoverage. They are generally regarded as undesirable for the reasons explained above.

It is not obvious how to incorporate the above desiderata on interval length into an algorithm that would be useful for choosing between different methods for setting limits.

5.1. *Two-sided intervals.* An alternative to giving upper limits is to quote two-sided intervals. For example, a 68% confidence interval for the mass of the top quark might be 169 to 173 GeV/$c^2$, as opposed to its 90% upper limit being 174 GeV/$c^2$. Most of the difficulties and ambiguities mentioned above apply in this case too, together with some extra possibilities. Thus, while it is clear which of two possible upper limits is tighter, this is not necessarily so for two-sided intervals, where which is shorter may be metric dependent; the first of two intervals for a particle's lifetime $\tau$ may be shorter, but the second may be shorter when the ranges are quoted for decay rate ($= 1/\tau$). Also, there is more scope for choice of ordering rule for the frequentist Neyman construction, or for choosing the interval from the Bayesian posterior probability density.[7]

It has been pointed out by Feldman and Cousins ([1998](#)) that an apparently innocuous procedure for choosing what result to quote may lead to undercoverage. Many physicists would quote an upper limit on any possible signal if their observation was not more than three standard deviations above the expected background, but a two-sided interval if their result was above this. With each type of interval constructed to give 90% coverage, there are some values of the parameter for which the coverage for this mixed procedure drops to 85%; Feldman and Cousins refer to this as "flip-flop." They circumvent the problem by using a "unified" approach, in which the method automatically yields upper limits for small values of the data, but two-sided intervals for larger measurements, while maintaining correct coverage for all possible true values of the signal.

5.2. *Sensitivity.* We have already mentioned the idea of quoting the sensitivity of a procedure, as well as the actual upper limit as derived from the observed data.[8] For upper limits or for uncertainties on measurements, this can be defined as the median value that would be obtained if the procedure was repeated a large number of times. Using the median is preferable to the mean because (a) it is metric independent (i.e., the median lifetime upper limit would be the reciprocal of the median decay rate lower limit); and (b) it is much less sensitive to a few anomalously large upper limits or error estimates.

Punzi ([2003](#)) has drawn attention to the fact that this choice of definition for sensitivity has some undesirable features. Thus, minimizing the median upper limit for a search provides a different optimization from maximizing the median number of standard deviations for the significance of a discovery. Also, there is only a 50% chance of achieving the median result or better. Instead, for pre-defined levels $\alpha$ and $CL$, Punzi determines at what signal strength there is a probability of at least $CL$ for establishing a discovery at a significance level $\alpha$. This is what he quotes as the sensitivity, and is the signal strength at which we are sure to be able to claim a discovery or to exclude its existence. Below this, the presence or otherwise of a signal makes too little difference, and we may remain uncertain (see Figure [2](#)).

5.3. $CL_s$. This is a technique [Read ([2000](#), [2004](#))] which is used for situations in which a discovery is not made, and instead various parameter values are excluded. For example, the Standard Model Higgs boson is such that, even before it is discovered, everything about it is well defined by theory except for its mass. The rate at which it is produced in a given experiment does depend on its mass. The failure to observe it can be converted into a mass range for the Higgs which is excluded (at some confidence level).

---

[7]A Bayesian statistician would be happy with the posterior as the final result. Particle Physicists like to quote an interval as a convenient summary.

[8]The sensitivity on its own will not do, because it is independent of the data.

Figure 3 illustrates the expected distributions for some suitably chosen test statistic under two different hypotheses: the null $H_0$ in which there is only standard known physics, and $H_1$ which also includes some specific new particle, such as the Higgs boson. In the simplest case, the statistic could be simply the observed number of events $n$ in some selected region. In Figure 3(c), the new particle is produced prolifically, and an experimental observation of $n$ should fall in one peak or the other, and easily distinguish between the two hypotheses. In contrast, Figure 3(a) corresponds to very weak production of the new particle and it is almost impossible to know whether the new particle is being produced or not. The conventional method of claiming new particle production would be if $n$ fell well above the main peak of the $H_0$ distribution; typically a $p_0$ value corresponding to $5\sigma$ would be required. In a similar way, new particle production would be excluded if $n$ were below the main part of the $H_1$ distribution. Typically, a 95% exclusion region would be chosen (i.e., $1 - p_1 \leq 0.05$). The $CL_s$ method aims to provide protection against a downward fluctuation of $n$ in Figure 3(a), resulting in a claim of exclusion in a situation where the experiment has no sensitivity to the production of the new particle; this could happen in 5% of experiments.

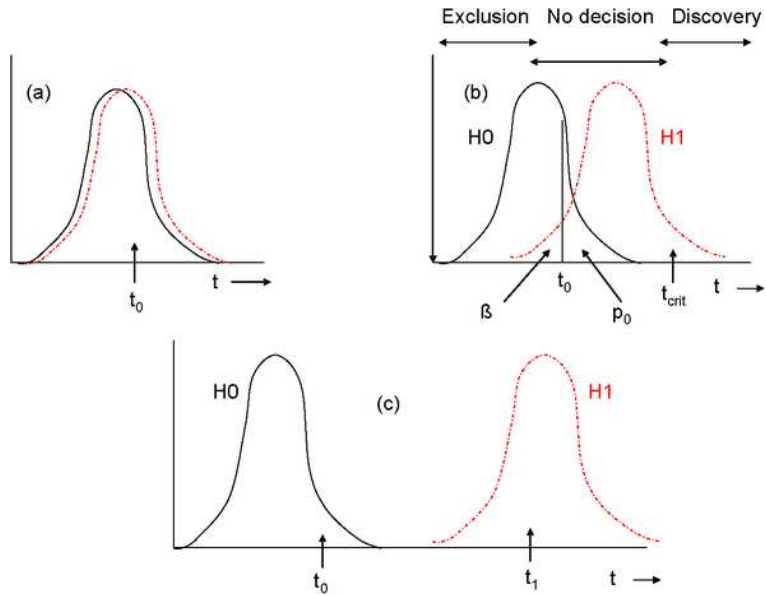

FIG. 2. *Punzi definition of sensitivity. Expected distributions for a statistic t (which in simple cases could be simply the observed number of events n), for $H_0$ = background only (solid curves) and for $H_1$ = background plus signal (dashed curves). In (a), the signal strength is very weak, and it is impossible to choose between $H_0$ and $H_1$. As shown in (b), which is for moderate signal strength, $p_0$ is the probability according to $H_0$ of t being equal to or larger than the observed $t_0$. To claim a discovery, $p_0$ should be smaller than some pre-set level $\alpha$, usually taken to correspond to $5\sigma$; $t_{crit}$ is the minimum value of t for this to be so. Then the power function $1 - \beta$ [equivalent to $p_1$ in Figure 3(b)] is the probability according to the alternative hypothesis that t will exceed $t_{crit}$. According to Punzi, the sensitivity should be defined as the expected production strength of the signal such that $1 - \beta$ exceeds another predefined level CL, for example, 95%. The exclusion region in (b) corresponds to $t_0$ in the 5% lower tail of $H_1$, while the discovery region has $t_0$ in the $5\sigma$ upper tail of $H_0$; there is a "No decision" region in between, as the signal strength in (b) is below the sensitivity value. The sensitivity is thus the signal strength above which there is a 95% chance of making a $5\sigma$ discovery. That is, the distributions for $H_0$ and $H_1$ are sufficiently separated that, apart possibly for the $5\sigma$ upper tail of $H_0$ and the 5% lower tail of $H_1$, they do not overlap. In (c) the signal strength is so large that there is no ambiguity in choosing between the hypotheses.*

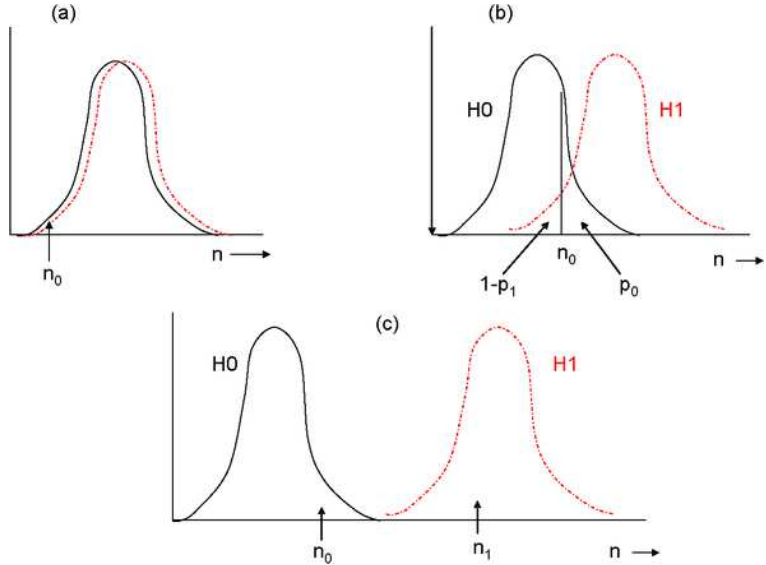

FIG. 3.   *The $CL_s$ method. The expected distributions for a data statistic $n$ are shown: (i) for the null hypothesis $H_0$ of background only (solid curve); and (ii) for $H_1$ (dashed curve), where there is also some exciting new physics, which tends to result in larger $n$. In (b), the tail areas of $H_0$ above the observed $n_0$ and of $H_1$ below $n_0$ are indicated by arrows; they correspond to probabilities $p_0$ and $1 - p_1$ respectively. Figure (c) shows a situation where the new physics is strongly produced, and $H_0$ and $H_1$ are well separated. Thus, $n_0$ would result in $H_1$ being excluded, while $n_1$ would be taken as evidence in favour of new physics. In (a), production is very weak, and the $H_0$ and $H_1$ curves are barely distinguishable. In order to protect against a downward fluctuation (statistic = $n_0$) in a situation like (a) resulting in an exclusion of $H_1$ when the curves are essentially identical, $CL_s$ is defined as $(1 - p_1)/(1 - p_0)$.*

It achieves this by defining[9]

(1)                         $$CL_s = (1-p_1)/(1-p_0),$$

and requiring $CL_s$ to be below 0.05. From the definition, it is clear that $CL_s$ cannot be smaller than $1-p_1$, and hence is a conservative version of the frequentist quantity $1-p_1$. It tends to $1-p_1$ when $n$ lies above the $H_0$ distribution, and to unity when $H_0$ and $H_1$ are very similar.

Statisticians may find $CL_s$, which is the ratio of two $p$-values, to be lacking in formal justification. Its appeal to Particle Physicists is the protection it provides against excluding particles from data which have no sensitivity to them. We thus regard it as a conservative frequentist approach.

5.4. *Nuisance parameters.* For calculating upper limits in the simple counting experiment described in Section 5, the nuisance parameters arise from the uncertainties in the background rate $b$ and the acceptance $\epsilon$. These uncertainties are usually quoted as $\sigma_b$ and $\sigma_\epsilon$ (e.g., $b = 3.1 \pm 0.5$), and the question arises of what these errors mean. Sometimes they encapsulate the results of a subsidiary measurement, performed to estimate $b$ or $\epsilon$, and then they would express the width of the Bayesian posterior or of the frequentist interval obtained for the nuisance parameters. However, in many situations, the errors may be based on a series of subsidiary measurements; they may involve Monte Carlo simulations, which have systematic uncertainties (e.g., related to how well the simulation describes the real data) as well as statistical errors; or they may reflect uncertainties or ambiguities in theoretical calculations required to derive $b$ and/or $\epsilon$. In the absence of further information the posterior is often assumed to be a Gaussian, usually truncated so as to exclude unphysical (e.g., negative) values. This may be at best only approximately true, and deviations are likely to be most serious in the tails of the distribution.

There are many methods for incorporating nuisance parameters in upper limit calculations. These include:

- *Profile likelihood.* The likelihood, based on the data from the main and from the subsidiary measurements, is a function of the parameter of interest $s$ and of the nuisance parameters. The profile likelihood $L_{\text{prof}}(s)$ is simply the full likelihood $L(s, b_{\text{best}}(s), \epsilon_{\text{best}}(s))$, evaluated at the values of the nuisance parameters that maximize the likelihood at each $s$. Then the profile likelihood is simply used to extract the limits on $s$, much as the ordinary likelihood could be used for the case when there are no nuisance parameters.

---

[9]Given the fact that $CL_s$ is essentially the ratio of two $p$-values, the choice of symbol $CL_s$ (standing for "confidence level of signal") is confusing.

Rolke et al. (2005) have studied the behavior of the profile likelihood method for limits. Heinrich (2003a) had shown that the likelihood approach for estimating a Poisson parameter (in the absence of both background and of nuisance parameters) can have poor coverage at low values of the Poisson parameter. However, the profile likelihood seems to do better, probably because the nuisance parameters have the effect of smoothing away the fluctuating coverage observed by Heinrich.

- *Full Bayes.* When there is a subsidiary measurement, a prior is chosen for $b$ (or $\epsilon$), the data is used to extract the likelihood, and then Bayes' theorem is used to deduce the posterior for the nuisance parameter. This posterior from the subsidiary measurement is then used as the prior for the nuisance parameter in the main measurement (this prior could alternatively come from information other than a subsidiary experiment); together with the prior for $s$ and the likelihood for the main measurement, the overall joint posterior for $s$ and the nuisance parameter(s) is derived.[10] This is then integrated over the nuisance parameter(s) to determine the posterior for $s$, from which an upper limit can be derived. Numerical examples of upper limits can be found in Heinrich et al. (2004), where the method is discussed in detail. Thus, for precisely determined backgrounds, the effect of a 10% uncertainty in $\epsilon$ can be seen for various measured values of $n$ in Table 2. A plot of the coverage when the uncertainty in $\epsilon$ is 20% is reproduced in Figure 4.

  It is not universally appreciated that the choice for the main measurement of a truncated Gaussian prior for $\epsilon$ and an (improper) constant prior for nonnegative $s$ results in a posterior for $s$ which diverges. Thus, numerical estimates of the relevant integrals are meaningless. Another problem comes from the difficulty of choosing sensible multi-dimensional priors. Heinrich has pointed out the problems that can arise for the above Poisson counting experiment, when it is extended to deal with several data channels simultaneously [Heinrich (2005)].

- *Fully frequentist.* In principle, the fully frequentist approach to setting limits when provided with data from the main and from subsidiary measurements is straightforward: the Neyman construction is performed in the multidimensional space where the parameters are $s$ and the nuisance parameters, and the data are from all the relevant measurements. Then the region in parameter space for which the observed data were likely is projected onto the $s$-axis, to obtain the confidence region for $s$.

  In practice, there are severe difficulties in writing a program to do this in a reasonable amount of time. To date, the largest number of parameters used is three [Nicola and Signorelli (2002)]. Another problem is that,

---

[10]This is usually equivalent to starting with a prior for $s$ and the nuisance parameters, and the likelihood for the data from the main and the subsidiary experiments together, to obtain the joint posterior.

Table 2

*90% confidence level upper limits for the production rate s as a function of n,
the observed number of events*

| $n$ | $b = 0.0$ | $b = 3.0$ |
|---|---|---|
| 0 | 2.35 (2.30) | 2.35 (2.30) |
| 3 | 6.87 (6.68) | 4.46 (4.36) |
| 6 | 10.88 (10.53) | 7.80 (7.60) |
| 9 | 14.71 (14.21) | 11.56 (11.21) |
| 20 | 28.27 (27.05) | 25.05 (24.05) |

The Poisson parameter $\mu = \epsilon * s + b$, where the expected background $b$ is either
0.0 or 3.0, and is precisely known; and $\epsilon$, whose true values is 1.0, is estimated
in a subsidiary measurement with 10% accuracy. The numbers in brackets
are the corresponding upper limits when $\epsilon$ is known precisely. At large $n$,
the limits for $b = 3.0$ are 3 units lower than those for $b = 0.0$; the latter are
approximately $n + 1.28\sqrt{n}$ at large $n$. The effect of the uncertainty in $\epsilon$ is to
increase the limits, and by a larger amount at large $n$. For $n = 0$, the Bayesian
limits are independent of the expected background $b$.

unless a clever ordering rule is used for producing the acceptance region in
data space for fixed values of the parameters, the projection phase leads to
overcoverage, which can become larger as the number of nuisance parameters increases. Good ordering rules have been found for a version of the
Poisson counting experiment [Punzi (2005)], and for the ratio of Poisson

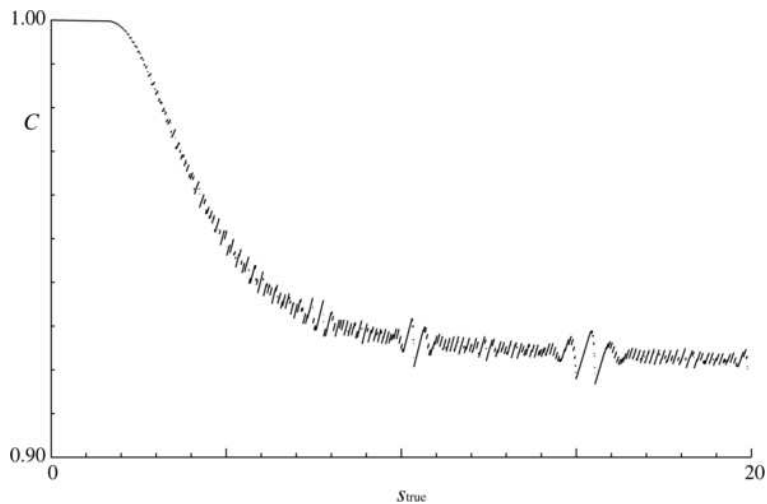

Fig. 4.   *The coverage $C$ for the estimated 90% confidence level upper limit as a function
of the true parameter $s_{\text{true}}$. The background $b = 3.0$ is assumed to be known exactly, while
the subsidiary measurement for $\epsilon$ gives a 20% accuracy. The discontinuities are a result of
the discrete (integer) nature of the measurements. There appears to be no undercoverage.*

means [Cousins ([1998](#))], where the confidence intervals are tighter than those obtained by conditioning on the sum of the numbers of counts in the two observations.

For the fully frequentist method, it is guaranteed that there will be no undercoverage for any combination of parameter true values. This is not so for any other method, and so most Particle Physicists would like assurance that the technique used does indeed provide reasonable coverage, at least for $s$. There is usually lively debate between frequentist and Bayesians as to whether coverage is desirable for all values of the nuisance parameter(s), or whether one should be happy with no or little undercoverage when experiments are averaged over the nuisance parameter true values.

• *Mixed.* Because of the difficulty of performing a fully frequentist analysis in all but the simplest problems, an alternative approach [Cousins and Highland ([1992](#))] is to use Bayesian averaging over the nuisance parameters, but then to employ a frequentist approach for $s$. The hope is that for most experiments setting upper limits, the statistical errors on the data are relatively large and so, provided the uncertainties in the nuisance parameters are not too large, the effect of the systematics on the upper limits will be small, and hence an approximate method of dealing with them may be justified.

5.5. *Banff challenge.* Given the large number of techniques available for extracting upper limits from data, especially in the presence of nuisance parameters, it was decided at the Banff meeting [Reid, Linnemann and Lyons ([2006](#))] that it would be useful to compare the properties of the different approaches under comparable conditions. This led to the setting up of the "Banff Challenge," which consisted of providing common data sets for anyone to calculate their upper limits. This was organized by Joel Heinrich, who reported on the performance of the various methods at the PHYSTAT-LHC meeting [Heinrich ([2007](#))].

**6. Discovery issues.** Searches for new particles are an exciting endeavor, and will play an even bigger role with the start-up of the LHC at CERN, expected in 2008. The 2007 PHYSTAT Workshop at CERN [Prosper, Lyons and De Roeck ([2007](#))] was devoted to statistical issues that arise in discovery-oriented analyses.

6.1. *p-values.* In order to quantify the chance of the observed effect being due to an uninteresting statistical fluctuation, some statistic is chosen for the data. The simplest case would be the observed number $n_0$ of interesting events. Then the *p*-value is calculated, which is simply the probability that, given the expected background rate $b$ from known sources, the observed number of events would fluctuate up to $n_0$ or larger. A small value of $p$ indicates that the data are not very compatible with the theory (which may be because we do not understand our detector, rather than the theory being wrong).

Particle physicists usually convert $p$ into the number of standard deviations $\sigma$ of a Gaussian distribution, beyond which the one-sided tail area corresponds to $p$. Thus, $5\sigma$ corresponds to a $p$-value of $3*10^{-7}$. This is done simply because it provides a number which is easier to remember, and not because Gaussians are relevant for every situation.

Unfortunately, $p$-values are often misinterpreted as the probability of the theory being true, given the data. It sometimes helps colleagues clarify the difference between $p(A|B)$ and $p(B|A)$ by reminding them that the probability of being pregnant, given the fact that you are female, is considerably smaller than the probability of being female, given the fact that you are pregnant.

6.2. *Nuisance parameters.* The calculation of $p$-values is complicated in practice by the existence of nuisance parameters. For example, for the simple situation described above, there could be some uncertainty in the estimated background. Although pivots are not generally used, there are numerous ways of incorporating nuisance parameters. These include:

- Conditioning: In simple cases with a single nuisance parameter, it may be possible to condition on the sum of the number of counts in the main and the subsidiary experiments, and then to use the binomial distribution to obtain the $p$-value.
- Plug-in $p$-value: The best estimate of the nuisance parameters is used to calculate $p$.
- Prior predictive $p$-value: The $p$-values are averaged over the nuisance parameters, weighted by their prior distributions.
- Posterior predictive $p$-value: This time, the posterior distributions of the nuisance parameters are used for weighting.
- Supremum $p$-value: The largest $p$-value for any possible value of the nuisance parameter is used. This is likely to be useful only when the nuisance parameter is forced to be within some range; or when there is only a finite number of possible alternative theoretical interpretations.
- Confidence interval: A confidence region of size $1 - \gamma$ is used for the nuisance parameter(s), and then the adjusted $p$-value is $p_{max} + \gamma$, where $p_{max}$ is the largest $p$-value as the nuisance parameters are varied over their confidence region. Clearly, if it is desired to establish a discovery from $p$-values around $10^{-7}$ or smaller, then $\gamma$ should be chosen at least an order of magnitude below this.

The properties of these and other methods are compared by Demortier (2007), while Cramer (2007) has discussed some of them in the context of searches at the LHC, where the distributions in the tails of the probability distributions for data can be very relevant. Again, any experience of Statisticians about incorporating nuisance parameters could result in useful advice.

The role of systematic effects is likely to be more serious here than for upper limits discussed in Section 5.4. This is because in upper limit situations the number of events is usually small, and so statistical errors dominate. In contrast, discovery claims have $p$-values of $3 * 10^{-7}$ or smaller, and so tails of distributions are likely to be important.

6.3. *Why 5σ?*   Unfortunately the usually accepted ideal for claiming a discovery in Particle Physics is that $p$ should correspond to at least $5\sigma$. Statisticians almost invariably ask why we use such a stringent level. One answer is past experience: we have all too often seen interesting effects at the $3\sigma$ or $4\sigma$ level go away as more data are collected. Another is the multiple comparison problem, or "look elsewhere" effect. While the chance of obtaining a $5\sigma$ effect in one bin of a particular histogram is really small, it is to be remembered that histograms have many bins,[11] they could be plotted with different selection criteria and different binning,[12] and there are very many other histograms that were or could have been looked at in the course of the experiment.[13] Thus, the chance of a $5\sigma$ fluctuation occurring somewhere in the data is much larger than might at first appear. Finally, physicists subconsciously incorporate Bayes' priors in assessing how likely they feel that they have discovered something new, and hence, whether they should claim a discovery. Thus, in deciding between the possibilities of a new discovery or of an undetected systematic effect, our priors might favor the latter, and hence, strong evidence for discovery is required from the data.

It is not necessarily equitable to use a uniform standard for large general-purpose experiments and for small ones with a specific aim; or for looking for a process which is expected, as compared with a very speculative search. But physicists and journal editors do like a defined rule rather than a flexible criterion, so this bolsters the $5\sigma$ standard. The general attitude is that, in the absence of a case for special pleading, $5\sigma$ is a reasonable requirement. In any case, it is largely a semantic issue, in that physicists finding a $4.5\sigma$ effect would clearly report it, using judiciously chosen wording to describe the status of their observation.

---

[11]In calculating a $p$-value in such a case, it is very desirable to take into account the number of chances for a statistical fluctuation to occur anywhere in the histogram. At very least, it should be made clear what the basis of the calculated $p$-value is.

[12]If a blind analysis is performed, such decisions are made before looking at the data, and so this aspect of the "look elsewhere" effect is reduced.

[13]The extent to which other people's searches should be included in an allowance for the "look elsewhere" effect depends subtly on the implied question being addressed. Thus are we considering the chance of obtaining a statistical fluctuation in any of the analyses we have performed; or by anyone analysing data in our experiment; or by any Particle Physicist this year? Anyone observing a possible Higgs signal at the LHC would be very unhappy about having to reduce the significance of their result because of the statistical fluctuations that could occur in speculative searches performed elsewhere.

Statisticians also ask whether we really believe our models out into the extreme tails of the distributions. In general, this may be so—counting experiments are expected to follow Poisson distributions, with small corrections for possible long time-scale drifts in detector calibrations; and particle decays usually are described by exponential distributions in time. However, the situation is much less clear for nuisance parameters, where error estimates may be less rigorous, and their distribution is often assumed to be Gaussian (or truncated Gaussian) by default. The effect of these uncertainties on very small $p$-values needs to be investigated case-by-case.

We also have to remember that $p$-values merely test the null hypothesis. A more sensitive way to look for new physics is via the likelihood ratio or the differences in $\chi^2$ for the two hypotheses, that is, with and without the new effect. Thus, a very small $p$-value on its own is usually not enough to make a convincing case for discovery.

6.4. *Repetitions in time.* A typical experiment at a large accelerator may collect data over 10–15 years. The same search for a new effect will typically be repeated once or twice each year as more data is collected. Does this constitute another factor of $\approx 20$ in the number of opportunities for a statistical fluctuation to appear? Our reply is "No." If there had been a $6\sigma$ signal with half the data (which resulted in a claim for discovery), which then became only $3\sigma$ with more data, this would be grounds for downplaying the earlier discovery claim. Thus, at any time, there is only one set of data (everything) that is relevant.

6.5. *Combining p-values.* In looking for a given new effect, there may be several separate and uncorrelated analyses which are relevant. These could correspond to different decay possibilities for the new particle or different experiments looking for the same signal. Thus, if the $p$-values for the null hypothesis (i.e., no new physics) for the separate analyses were $10^{-6}$ and 0.1, what is the corresponding $p$-value for the pair of results[14]?

The unambiguous answer is that there is no unique recipe for combining them [CDF (2007), Cousins (2007)]. There is no single way of taking a uniform distribution in two variables, and finding a transformation $p_{\text{comb}}(p_1, p_2)$ that converts it into a uniform distribution of the single variable $p_{\text{comb}}$.

---

[14]Rather than combining $p$-values, it is of course better to use the complete sets of original data (if available) for obtaining the combined result.

Two popular recipes involve asking what is the probability that the smaller $p$-value will be $10^{-6}$ or smaller or that the product is below $p_1 * p_2 = 10^{-7}$. None of the possible methods has the property that in combining 3 $p$-values, the same answer is obtained if $p_1$ is first combined with $p_2$, and then the result is combined with $p_3$; or whether some different ordering is used. Clearly, it is important to decide what combination method should be used, without reference to the specific data.

6.6. *Peak above smooth background.* When comparing two hypotheses with our data, we can use the numerical values of the two $\chi^2$ quantities. For example, we may be fitting a smooth distribution by a power series, and wonder whether we need a quadratic term, or whether a linear expression would suffice. Alternatively, we may want to assess whether a mass spectrum favors the existence of a peak on top of a smooth background, as compared with just the smooth background. Qualitatively, if the extra term(s) are unnecessary, they will result in a relatively small reduction in $\chi^2$, while if they really are required, the reduction could be larger.

It is sometimes possible to be quantitative about the expected reduction when the extra terms are not needed [Wilks (1938)]. If we are in the asymptotic regime, and if the hypotheses are nested, and if the extra parameters of the larger hypothesis are defined under the smaller one, and in that case do not lie on the boundary of their allowed region, then the difference in $\chi^2$ should itself be distributed as a $\chi^2$, with the number of degrees of freedom equal to the number of extra parameters.

An example that satisfies this is provided by the different order polynomials. Provided we have a large amount of data, we expect the difference in $\chi^2$ to have one degree of freedom, so a value larger than around 5 would be unlikely.

A contrast is provided by a smooth background $C(x)$ compared with a background plus peak, $C(x) + A \exp[-0.5 * (x - x_0)^2 / \sigma^2]$. The extra parameters for the peak are its amplitude, position and width: $A$, $x_0$ and $\sigma$ respectively. Again, the hypotheses are nested, in that $C(x)$ is just a special case of the peak plus background, with $A = 0$. However, although $A$ is defined in the background only case, $x_0$ and $\sigma$ are not, as their values become completely irrelevant when $A = 0$. Furthermore, unless the peak plus background fit allows $A$ to be negative, zero is on the boundary of its allowed region. We thus should not expect the difference of the $\chi^2$ quantities itself to be distributed as a $\chi^2$ [Protassov et al. (2002), Demortier (2006)]. To assess the significance of a particular $\chi^2$ difference, this unfortunately means that we have to obtain its distribution ourselves, presumably by Monte Carlo. If we want to find out probabilities of statistical fluctuations at the $10^{-6}$ level, this requires a lot of simulation, and probably needs us to use something better than brute force.

Another example of comparing hypotheses by their $\chi^2$ values is given in Section 11.3.

The problem of nonstandard limiting distributions for $\chi^2$ tests has a substantial statistical literature [see, e.g., Self and Liang (1987) and Drton (2007)].

**7. Goodness-of-fit.** With sparse data, the unbinned likelihood method is a good one for estimating parameters of a model. In order to understand whether these estimates of the parameters are meaningful, we need to know whether the model provides an adequate description of the data. Unfortunately, as emphasised by Heinrich (2003b), maximum likelihood is often insensitive to whether or not the data agree with the model. It would be very useful to have a way of utilizing the unbinned likelihood so that it does provide a measure of the goodness-of-fit.

The standard method loved by Particle Physicists is $\chi^2$. This, however, is only applicable to binned data (i.e., in a one or more dimensional histogram). Furthermore, it loses its attractive feature that its distribution is model-independent when there are not enough data, which is likely to be so in the multi-dimensional case.

An alternative that is used for sparse one-dimensional data is the Kolmogorov–Smirnov (KS) approach or one of its variants. However, in the presence of fitted parameters, simulation is again required to determine the expected distribution of the KS-distance. Also because of the problem of how to order the data, it is not used by Particle Physicists in multi-dimensional situations.

Aslan and Zech (2004, 2005) have described a method that can be used with sparse multi-dimensional data.[15] It compares two separate sets of events, which could be data and simulation based on a theoretical model or two sets of data taken under slightly different conditions, etc. The first set of points are assigned positive electric charges, and the second set negative ones, and then the "electrostatic energy" of the system is calculated as $E = \Sigma\Sigma q_i * q_j * f(d_{ij})$, where the summation extends over all pairs of observations; $q_i$ is the charge of the $i$th observation; and $f(d_{ij})$ is a function of the distance $d_{ij}$ between observations $i$ and $j$. For real electrostatics in 3 dimensions, $f(d)$ is proportional to $1/d$, but here it can be chosen to give desirable behavior; Aslan and Zech favor $-\ln(d + \epsilon)$, where $\epsilon$ is a small constant to avoid problems as $d$ tends to zero. This method requires the choice of a metric for each of the observables, and it also needs simulation to determine the expected distribution of $E$ assuming the two distributions are identical. Aslan and Zech find that their method compares favorably with other approaches (e.g, $\chi^2$, KS and its variants, etc.) in rejecting alternative hypotheses in various one-dimensional problems.

---

[15]A similar approach can be found in the statistics literature [Cuadras, Fortiana and Oliva (1997, 2003)].

**8. Blind analyses.** These are becoming increasingly popular in Particle Physics, as a means of avoiding personal bias affecting the result. They involve keeping part of the data unseen by the physicists, until the data selection procedure and the analysis method have been completely defined, all correction procedures specified, etc.

The original suggestion to use a blind analysis for a Particle Physics experiment was due to Luis Alvarez. An experiment at Stanford had looked for quarks, by measuring the residual charge on small spheres that were levitated in a superconducting magnet. If a single free quark was present in a sphere, the residual charge would be a third or two-thirds of the electron's charge. Several of the balls tested indeed yielded such values. A potential problem was that large corrections had to be applied to the raw data in order to extract the final result for the charge. The suspicion was that maybe the experimenters were (subconsciously) applying corrections until the value turned out to be "satisfactory." The blind approach would involve the computer adding a random number to the raw value of the charge, which would then be corrected until the experimentalists were satisfied, and only then would the computer subtract the random number to reveal the final answer for that sphere.[16]

There are various methods of performing blind analyses [Klein and Roodman (2005)], most of which aim to allow the experimentalists to look at some of the real data, in order to perform checks that nothing is terribly wrong. Some of these are as follows:

- The computer adds a random number to the data, which is only subtracted after all corrections are applied. This was the method suggested by Alvarez.
- Use only Monte Carlo to define the procedure. This completely avoids the danger of allowing the data to determine the procedure to be used, but suffers from the drawback that the data cannot be compared with the Monte Carlo, to check that the latter is reasonable.
- Use only a fraction of the data for defining the procedure. Then this is held fixed for the remainder of the data. In principle, an optimization can be employed to determine the fraction to be kept open, but, in practice, this is often decided by choosing a semi-arbitrary time after which the future data is kept blind.
- The signal region is defined as a certain part of multi-dimensional space, and this is kept hidden, but all other regions, including those adjacent to the signal, are available for inspection.

---

[16]This suggestion was implemented, but in fact no subsequent results were published. The current consensus is that this "discovery" of free quarks is probably spurious.

- Keep the Monte Carlo parameters hidden. This is a technique used by the TWIST experiment in their high statistics precision determination of parameters associated with muon decay. The procedure involves comparing the data with various simulated sets, generated with a series of different parameter values. The data and the simulations are both visible, but the parameter values used to generate the simulations are kept hidden.
- Keep visible only a fraction of the contents of each bin of a histogram. This is used by the MINOS experiment searching for neutrino oscillations; these would affect the energy distribution of the observed events. By keeping visible different unknown fractions of the data in each bin, the energy spectral shape cannot be determined from the visible part of the data.

If several different groups within the same collaboration are performing similar analyses for extracting some specific parameter, then it is desirable to fix the procedure for selecting which result to present, or alternatively how to combine the separate results. This should be done before the results are seen, and is worth doing even if the individual analyses were not "blind."

A question that arises with blind analyses is whether it should be permitted to modify the analysis after the data had been unblinded. It is generally agreed that this should not be done… unless everyone would regard it as ridiculous not to do so. For example, if a search for rare events yielded 10 candidates over the course of a year's run, and it was found that all of these occurred on Sunday mornings at precisely 1:17 a.m., it would be prudent to do some further investigation before publishing. If "post-unblinding" modification of the procedure is performed, this should be made clear in any publication.

**9. Combining results.** A commonly used procedure is to combine $N$ different uncorrelated measurements $a_i \pm \sigma_i$ of the same physical quantity $a$. When the measurements are believed to be Gaussian distributed about the true value $a_{\text{true}}$, the well-known result is that the best estimate $a_{\text{best}} \pm \sigma_{\text{best}}$ is given by

$$(2) \qquad a_{\text{best}} = \Sigma a_i * w_i / \Sigma w_i, \qquad \sigma_{\text{best}} = 1/\sqrt{\Sigma w_i},$$

where the weights are defined as $w_i = 1/\sigma_i^2$. This is readily derived from minimizing with respect to $a$ a weighted sum of squared deviations[17]

$$(3) \qquad\qquad\qquad S(a) = \Sigma(a_i - a)^2/\sigma_i^2.$$

---

[17]A problem arises if the measurements are discrepant. If $S$ is much larger than $N-1$, then some serious problem exists, and it is probably unwise to combine the results. But for $S/(N-1)$ somewhat larger than unity, a commonly adopted procedure [Particle Data Group (2006)] is to scale up the uncertainty on the weighted average by the square root of this factor.

The extension to the case where the individual measurements are correlated (as is often the case for analyses using different techniques on the same data) is straightforward: $S$ becomes $\Sigma\Sigma(a_i - a) * H_{ij} * (a_j - a)$, where $H$ is the inverse error matrix.

There are, however, practical details that complicate its application. For example, in the above formula, the $\sigma_i$ are supposed to be the **true** accuracies of the measurements. Often, all that we have available are **estimates** of their values. Problems arise in situations where the error estimate depends on the measured value $a_i$. For example, in counting experiments with Poisson statistics, it is typical to set the error as the square root of the observed number. Then a downward fluctuation in the observation results in an overestimated weight, and $a_{\text{best}}$ is biassed downward. If instead the error is estimated as the square root of the expected number $a$, the combined result is biassed upward—the increased error reduces $S$ at larger $a$. A way round this difficulty has been suggested by Lyons, Martin and Saxon ([1990](#)).

Another problem arises when the individual measurements are very correlated. When the correlation coefficient of two uncertainties is larger than $\sigma_1/\sigma_2$ (where $\sigma_1$ is the smaller error), $a_{\text{best}}$ lies outside the range of the two measurements. As the correlation coefficient tends to $+1$, the extrapolation becomes larger, and is very sensitive to the exact value assumed for the correlation coefficient. The situation is aggravated by the fact that $\sigma_{\text{best}}$ tends to zero. This is usually dealt with by selecting one of the two analyses, rather than trying to combine them.

Another extension of this procedure is for combining $N$ pairs of correlated measurements (e.g., the gradient and intercept of a straight line fit to several sets of data). The prescription to be adopted for scaling the errors when the individual measurements are somewhat discrepant has complications.

**10. Accuracy of answer.** Sometimes a result appears to be more accurate than is justified. This can arise when an upper limit is much lower than the sensitivity of the procedure (e.g., when the observed number of events in a counting experiment is smaller than the expected background) or when by chance individual observations happen to lie close to each other. This can cause problems in deciding which measurement is "better." This can be relevant in choosing which of several competing analyses on the same data to quote as the result of the experiment; or in combining different results (see previous section).

In the former situation, if the estimated error increases with the estimated value, choosing the result with the smallest **estimated** error can produce a downward bias. On the other hand, using the smallest **expected** error can cause us to ignore an analysis which had a particularly favorable statistical fluctuation, which produced a result that was genuinely more precise than expected.[18] How to deal with this situation in general is an open question. It has features in common with the problem of measuring a voltage by choosing at random a voltmeter from a cupboard containing meters of different sensitivities [Cox (1958)].

**11. Recent improvements in understanding.** In this section we list a few of the issues on which Particle Physicists have recently improved their understanding of statistical issues. To those can be added a few already discussed above (see Section 6.6 and the remarks about unbinned likelihoods in the first paragraph of Section 7).

11.1. *Number of degrees of freedom.* If we construct the weighted sum of squares $S$ between a predicted theoretical curve and some data in the form of a histogram, provided the Poisson distribution of the data can be approximated by a Gaussian (and the theory is correct, the data are unbiassed, the error estimates are correct, etc.), **asymptotically**[19] $S$ will be distributed as $\chi^2$ with the number of degrees of freedom $\nu = n - f$, where $n$ is the number of data points and $f$ is the number of free parameters whose values are determined in the fit.

The relevance of the asymptotic requirement can be seen by imagining fitting a more or less flat distribution by the expression $N(1 + 10^{-6} \cos(x - x_0))$, where the free parameters are the normalization $N$ and the phase $x_0$. It is clear that, although $x_0$ is left free in the fit, because of the $10^{-6}$ factor, it will have a negligible effect on the fitted curve, and hence will not result in the typical reduction in $S$ associated with having an extra free parameter. Of course, with an enormous amount of data, we would have sensitivity to $x_0$, and so asymptotically it does reduce $\nu$ by one unit, but not for smaller amounts of data.

Another example involves the search for neutrino oscillations. The neutrino energy spectrum is fitted by a survival probability $P$ of the form

$$(4) \qquad P = 1 - A \sin^2(C * \Delta m^2),$$

where $C$ is a known function of the neutrino energy and the length of its flight path, $A$ is a parameter which depends on the neutrino mixing angle, and $\Delta m^2$ is the difference in mass squared of the relevant neutrino species. For small values of $C * \Delta m^2$,

$$(5) \qquad P \approx 1 - A(C * \Delta m^2)^2.$$

---

[18]For example, the ALEPH experiment at LEP produced a tighter-than-expected upper limit on the mass of $\nu_\tau$ because they happened to observe a decay configuration producing $\nu_\tau$ which was particularly sensitive for determining its mass.

[19]The examples in this section are independent of the requirement that we need enough events for the Poisson distribution to be well approximated by a Gaussian.

Thus, the survival probability depends only on the two parameters in the combination $A (\Delta m^2)^2$. Because this combination is all that we can hope to determine, we effectively have only one free parameter rather than two. Of course, an enormous amount of data can manage to distinguish between $\sin(C * \Delta m^2)$ and $C * \Delta m^2$, and so asymptotically we have two free parameters as expected.

It would be useful to have some indication of when data are near enough to asymptopia, so as to avoid the necessity for Monte Carlo calculations of the expected distribution of $S$.

11.2. $\Delta(\ln L) = 0.5$ *rule.* In the maximum likelihood approach to parameter determination, the best value $\lambda_0$ of a parameter is determined by finding where the likelihood maximizes; and its error $\sigma_\lambda$ is estimated by finding how much the parameter must be changed in order for the logarithm of the likelihood to decrease by 0.5 as compared with the maximum.[20] From a frequentist viewpoint, this should ideally result in the range from $\lambda_0 - \sigma_\lambda$ to $\lambda_0 + \sigma_\lambda$ having 68% coverage.

If the measurement is distributed about the true value as a Gaussian with a constant width, then exact coverage is obtained, but in general this is not so. For example, Heinrich (2003a) has investigated the properties of the likelihood approach to estimate $\mu$, the mean of a Poisson, when $n_{\text{obs}}$ events are observed. Because $n_{\text{obs}}$ is a discrete variable, the coverage is a discontinuous function of $\mu$, and varies from 100% at $\mu = 0$ down to 30% at $\mu \approx 0.5$.[21]

11.3. *Comparing two hypotheses via* $\chi^2$. Assume we have a histogram with 100 bins, and that we are using a $\chi^2$ method for fitting it with a function with one free parameter. We expect to obtain a $\chi^2$ value of $99 \pm 14$. Thus, if $p_0$, the best value of the parameter, yields a $\chi^2$ of 85, we would regard that as very satisfactory. However, a theoretical colleague has a model which predicts that the parameter should have a different value $p_1$, and wants to know what the data has to say about that. We test this by calculating the $\chi^2$ for that $p_1$ and obtain a value of 110. We appear to have two contradictory conclusions:

- $p_1$ is satisfactory: This is based on the fact that the relevant $\chi^2$ of 110 is well within the expected range of $99 \pm 14$.

---

[20]If there is more than just one parameter, the likelihood must be remaximized with respect to all the other parameters when looking for the $\Delta(lnL) = 0.5$ points.

[21]It is of course not surprising that methods that are expected to have good asymptotic behavior may not display optimal properties for $\mu \approx 0$.

- $p_1$ is ruled out: The uncertainty on $p$ is estimated by seeing how much it must change from its optimum value in order to make $\chi^2$ increase by 1 unit. For this data, $\chi^2(p_1)$ is **25** units larger than $\chi^2(p_0)$, and so, assuming that the behavior of $\chi^2$ in the neighborhood of the minimum is parabolic, $p_1$ is ruled out at the 5 standard deviation level.

Unfortunately, many physicists, over-impressed by the fact that $\chi^2(p_1)$ appears to be satisfactory, are reluctant to accept that $p_0$ is strongly favored by the data.

A similar argument applies to comparing a given set of data with 2 separate hypotheses, for example, fitting a histogram with an exponential or a straight line. Again the **difference** between the $\chi^2$ quantities provides better discrimination between the hypotheses than do the **individual** $\chi^2$ [Lyons (1999)].

There are of course other ways of comparing two hypotheses e.g. likelihood ratio, Bayes factor, Bayesian information criterion, etc. Trotta (2008) has discussed their application in cosmology.

**12. Conclusions.**  It is clear that there are many practical issues to be resolved in Particle Physics. Some of these may be of interest to Statisticians. With analyses becoming more and more complex, we would welcome more active involvement that would lead to improved analyses of our data. Any suggestions regarding improvements in the approaches outlined in this review would also be appreciated.

**Acknowledgments.**  I wish to acknowledge the patience and expertise of David Cox, Brad Efron and Michael Stein and also of other Statisticians too numerous to list, in explaining statistical issues to me; the ones who have contributed to the PHYSTAT meetings have been particularly helpful. My understanding of the practical application of statistical techniques has improved considerably as a result of discussions with many experimental Particle Physics colleagues. I especially want to thank the members of the CDF Statistics Committee and Bob Cousins. To all of you, I am most grateful.

### SUPPLEMENTARY MATERIAL

**Appendix: Glossary of Particle Physics terms** (DOI: 10.1214/08-AOAS163SUPP; pdf).

PARTICLE PHYSICS
DENYS WILKINSON BUILDING
UNIVERSITY OF OXFORD
KEBLE ROAD
OXFORD OX1 3RH
UNITED KINGDOM
E-MAIL: l.lyons@physics.ox.ac.uk